\documentstyle[12pt,epsf]{article}

\title{
\begin{flushright}
{\large Yaroslavl State University\\
        Preprint YARU-HE-96/02} \\[10mm]
\end{flushright}
       {\LARGE\bf The radiative decay of the massive neutrino} \\
       {\LARGE\bf in the external electromagnetic fields}}

\author{{\Large\bf A.A.~Gvozdev, N.V.~Mikheev} \\[2mm]
        {\large\it
             Division of Theoretical Physics, Department of Physics,} \\
        {\large\it
             Yaroslavl State University, Yaroslavl 150000, Russia} \\[4mm]
        {\Large\bf and} \\[4mm]
        {\Large\bf L.A.~Vassilevskaya} \\[2mm]
        {\large\it
             Moscow Lomonosov University, V-952, Moscow 117234, Russia} \\
        {\large\it E-mail: vassilev@yars.free.net, vasilevs@vitep5.itep.ru}}

\date{}

\begin{document}

\maketitle

\begin{abstract}

The radiative decay of the massive neutrino $\nu_i \rightarrow \nu_j
\gamma$ is investigated in the framework of the Standard Model
in external electromagnetic fields of various
configurations: constant crossed field, constant uniform magnetic field,
plane monochromatic wave's field. The effect of significant enhancement
of the neutrino decay probability by the external field
(electromagnetic catalysis) is discussed. An especially strong enhancement
occurs in the case of the ultrarelativistic neutrino decay, since in this
case the decay probability does not contain suppression caused by
the smallness of the decaying neutrino mass. The
ultrarelativistic neutrino decay catalysis is significant even in a
relatively weak external field ($F/F_e \ll 1$, where $F_e$ is the critical
Schwinger value). The expression for the photon splitting probability
into the neutrino pair $\gamma \rightarrow \nu_i \bar\nu_j$ in
the wave field is given. The estimations of a number of gamma-quanta
produced in a volume filled with an electromagnetic field and the neutrino
lifetime in a strong magnetic field are presented.

\end{abstract}

%\eqnumsection

\vspace*{10mm}

\centerline{To be published in Physical Review, V.~{\bf D54}, N~7. }

\thispagestyle{empty}

\newpage

\large

\section{Introduction}

For quite a long time we have seen intensive theoretical studies of
flavour-changing processes caused by the phenomenon of fermion
mixing.
The  description  of  this
phenomenon in  the quark sector goes back to the pioneer work  by
Cabibbo~\cite{C}
and presently is put into practice  by  introducing  the
unitary $3 \times 3$ matrix $V_{ij}$ (the so  called
Cabibbo-Kobayashi-Maskawa matrix~\cite{KM}.
It should be noted that qualitative progress has been
observed  in   experimental   investigation   of   quark   mixing
parameters. Except a detailed examination  of  the "old" mixing
angles related  to  the  first  two  quark  generations,  important
information  has  been  obtained  from  the  decays  of $b$-quark
containing  particles (ARGUS Collaboration~\cite{AR},
CLEO Col\-la\-bo\-ra\-tion~\cite{CL}). On  the  other
hand, thus far there is no experimental evidence in favor of  the
analogous mixing phenomenon in the lepton  sector.  This  can  be
accounted for in a natural way by the fact that, because  of  the
insufficiently high precision achieved in experimental studies of
neutrino-involving processes, the neutrino mass spectrum  appears
degenerate  (the  neutrinos  manifest  themselves  as  massless
particles~\cite{RPP}). The neutrino mass spectrum being degenerate,
lepton mixing is known to be purely formal and  unobservable.  At
the same time, with massive neutrino the absence of lepton mixing
seems unnatural and is virtually incompatible  with  attempts  to
somehow extend the standard model. Notice that lepton mixing
may lead to some interesting  physical
phenomena such as:

\noindent 1) charged lepton radiative decays with lepton number violation
of type  $\mu \rightarrow e \gamma$,
$\mu \rightarrow 3e$~\cite{Pet,SCH},
$\mu \rightarrow e \gamma \gamma$~\cite{GMV1}.

\noindent 2) neutrino radiative decays $\nu_i \rightarrow \nu_j \gamma$~
\cite{Pet},
$\nu_i \rightarrow \nu_j \gamma \gamma$ \cite{N};

\noindent 3) neutrino oscillations \cite{Pon},

\noindent 4) the possible effect of massive neutrino mixing on the
spectrum of $\beta$ -- decay-produced  electrons  \cite{ShK}.

Even such a short review of lepton mixing effects shows that
most of these are associated with the massive neutrino. Nowadays,
the physics of the massive  neutrino  is  becoming  a  vigorously
growing and prospective line of investigation at the junction  of
elementary particles physics, astrophysics and cosmology. It will
suffice to mention the well known problem of the  solar  neutrino
\cite{SN}
and  the  possibility  of  solving  it  (the  mechanism  of
resonance  enhancement  of  neutrino  oscillations  in  substance
\cite{MSW}),
the effect of the massive neutrino radiative decay on  the
spectrum of the relic radiation \cite{AV} and so on.
The above-mentioned way  of  solving  the  solar
neutrino problem using the
MSW mechanism\footnote{About the current situation around the
solar neutrino problem see, for example, ref's~\cite{HL}}
shows that the  the
massive neutrino's properties are  sensitive  to  the  medium  it
propagates  through.  Substance  is  usually  considered  as
medium. We note, however, that  medium can also be represented
by an  external  electromagnetic  field, which  can  significantly
influence both the properties of the massive neutrino itself \cite{MP}
and the process of its decay \cite{GMV2}
and even  induce  novel  lepton transitions  with flavour violation
$\nu_i \leftrightarrow \nu_j$ ($i \neq j$) \cite{BTV},
forbidden in vacuum. In our preliminary communication \cite{GMV2}
we have  pointed
out the probability  of  the  massive  neutrino  radiative  decay
$\nu_i \rightarrow \nu_j \gamma$ ($i \neq j$) being considerably
enhanced in  a  constant  uniform
magnetic field. Such an enhancing influence of an external  field
can be illustrated with the straightforward example of a neutrino
radiative decay in a weak (as compared with the  Schwinger  value
$F_e = m^{2}_{e}/e \simeq 4.41 \cdot 10^{13} \, G$) electromagnetic
field. To  this  end  we
use the amplitude of the  Compton-like  process
$\nu_i \gamma^{\ast} \rightarrow \nu_j \gamma^{\ast}$  with
virtual photons \cite{KuM}, which, in particular, allows obtaining  the
first term of  the  expansion  of  the  radiative  decay
$\nu_i \rightarrow \nu_j \gamma$
amplitude in a weak external field. In  the  expression  for  the
amplitude of the  process $\nu_i (p_1) + \gamma^\ast (q_1)
\rightarrow  \nu_j (p_2) + \gamma^\ast(q_2)$
it is sufficient to consider $\gamma (q_2)$ as a real photon,
and to replace the field tensor of the virtual photon
$\gamma^\ast (q_1)$ by the  Fourier  image
of the external electromagnetic field tensor. Below we shall give
the expression obtained in  this  way  for  the  radiative  decay
amplitude in the  simplest  case  of  a  uniform  electromagnetic
field, in which the decay kinematics $p_1 + 0 = p_2 + q_2$ is the  same
as in vacuum. The external-field-induced contribution $\Delta {\cal M}$
to  the amplitude of the decay $\nu_i \rightarrow \nu_j \gamma$ can be
represented in the following form:

\begin{equation}
\Delta {\cal M} \simeq  {e \over 48 \pi^2} \, {G_{F} \over \sqrt{2}} \,
(jq) \, (F \tilde{f}^{*}(q)) \left < {1 \over F_{\ell}} \right > .
\label{eq:DM}
\end{equation}

\noindent where
$j_\mu = \bar \nu_j (p_2) \gamma_\mu (1 + \gamma_5) \nu_i(p_1)$,
$i,j=1,2,3$   enumerate   the
definite mass neutrino species,
$p_{1}$, $p_{2}$, $q$ are the  four-momenta  of
the initial and final neutrinos and the photon, respectively,
$F_{\mu \nu }$ is the external uniform electromagnetic field tensor,
$\tilde{f}_{\alpha \beta }(q) = \epsilon _{\alpha \beta \mu \nu}
q_{\mu} \epsilon_{\nu }(q)$,
$\epsilon _{\nu }(q)$  is  the  polarization  four-vector of the photon,
$F_\ell = m^2_\ell / e$ is the critical  value  of  the  strength
of the electromagnetic field for the charged lepton with the mass
$m_{\ell}$. We have introduced the following designation :

\begin{equation}
< A (m_\ell) > = \sum^{}_{\ell = e, \mu ,\tau} K_{i \ell} K^{*}_{j \ell}
\, A (m_\ell),    \label{eq:D<>}
\end{equation}

\noindent where  $K_{i \ell}$   ($\ell = e, \mu ,\tau$)   is   the
lepton   mixing   matrix   of Kobayashi-Maskawa type.
For the sake of comparison we  write  the
known expression for the  amplitude  of  the  neutrino  radiative
decay in vacuum \cite{Pet}
which can be represented as follows:

\begin{equation}
{\cal M}_{0} \simeq  -i \;{3 e \over 32 \pi^2} \, {G_{F} \over \sqrt 2} \,
(j \tilde{f}^{*}(q) p) \left < {m^{2}_{\ell} \over m^{2}_{W}} \right > ,
\label{eq:M0}
\end{equation}

\noindent where $p = p_{1} + p_{2}$,  $m_{\ell}$, $m_{W}$ are the masses
of  the  virtual  lepton  and
$W$-boson, respectively. In  analyzing  the  amplitudes (\ref{eq:DM})
and (\ref{eq:M0})
in the case of the neutrino decay at rest, it is  necessary
to take account of $p_{1}$, $p_{2}$, $q$, $j$ being of order of the
mass  of  the decaying neutrino $m_{\nu}$.
In  this  case, the  expressions  for  the
amplitudes (\ref{eq:DM}), (\ref{eq:M0})
can be easily estimated (it is sufficient
to  allow  for  the  order  of   the   dimensional   quantities):

\begin{eqnarray}
\label{eq:ADM}
\Delta {\cal M} & \sim & G_{F} m^{3}_{\nu} ( F / F_{e}), \\
\label{eq:AM0}
{\cal M}_{0} & \sim & G_{F} m^{3}_{\nu} (m_{\tau} / m_{W})^{2}.
\end{eqnarray}

\noindent It  follows  here  from   that,   given   the
condition

\begin{equation}
(F / F_{e})^{2} \gg (m_{\tau} / m_{W})^{4}  \label{eq:C}
\end{equation}

\noindent (here $F$ stands for the strengths of the magnetic ($B$) and
electric ($E$) fields), the probability of the decay
$\nu_i \rightarrow \nu_j \gamma$ in  an  external
field is much greater than that in vacuum, even for a  relatively
weak electromagnetic field ($10^{-3} \ll  F/F_e \ll 1$). The  catalyzing
effect of an external field becomes even more substantial in  the
case of the ultrarelativistic neutrino decay  ($E_\nu \gg m_\nu$).
With  the amplitudes (\ref{eq:DM}), (\ref{eq:M0}),
being Lorentz-invariant, the analysis can
be conveniently carried out in the rest  frame  of  the  decaying
neutrino. In this case the the electromagnetic field in Eq.~(\ref{eq:ADM})
is obtained by the Lorentz  transformation  from  the  laboratory
frame, in which the external field $F$ is given, to the rest  frame
of the decaying neutrino:

\begin{equation}
F' \sim { E_\nu \over m_\nu} F \gg  F.  \label{eq:UF}
\end{equation}

\noindent Comparing the expressions (\ref{eq:ADM}) and (\ref{eq:AM0}),
in view of  (\ref{eq:UF}),   we
notice that the catalyzing effect of the external  field  becomes
appreciable under a much weaker condition, as compared to (\ref{eq:C}):

\begin{equation}
{(p_1 F F p_1) \over m^2_\nu F^2_e} \gg
\left ( {m_{\tau} \over m_{W}} \right )^4.    \label{eq:UE}
\end{equation}

In  this  case  the  ratio  between  the  probabilities  of   the
ultrarelativistic  neutrino  decay $w^{(F)}$   and   the   decay   in
vacuum $w^{(0)}$  is  of  the  order:

\begin{equation}
{w^{(F)} \over w^{(0)}} \sim
\left ( {F \over F_{e}} \right )^2
\left ( {E_{\nu } \over m_{\nu}} \right )^2
\left ( {m_{W} \over m_{\tau}} \right )^4 \gg 1.  \label{eq:RW}
\end{equation}

\noindent The expression (\ref{eq:RW}) shows that in the
ultrarelativistic
neutrino decay the enhancement is mainly due to a decrease  in the decay
probability suppression by the smallness  of  the  neutrino  mass
($w^{(F)} \sim m^{4}_{\nu}$,
$w^{(0)} \sim m^{5}_{\nu} (m_{\nu} / E_{\nu})$).
It is  natural  to  expect  that  in
taking the account of further  terms  in  the  expansion  of  the
amplitude of the radiative  decay $\nu_i \rightarrow \nu_j \gamma$
with  respect  to  the
external field, the suppression  mentioned  above  can  be  fully
canceled.  All  this  makes  it  interesting  to  calculate   the
amplitude with the  external  electromagnetic  field  taken  into
account exactly. An expression thus obtained will be valid in the
case of the  neutrino  radiative  decay $\nu_i \rightarrow \nu_j \gamma$
in  an  external
electromagnetic field, which has not to be weak  as  against  the
Schwinger value $F_e$.

\section{The crossed field}

At  present the  experimentally  accessible  strengths  of
electromagnetic  fields  are  significantly  below  the  critical
strength ($F/F_e \ll  1$, $F=B,{\cal E}$,
$F_{e}= m^{2}_{e}/e \simeq 4.41 \cdot 10^{13} \, G$). Because  of
this  ,  field-induced  effects  are  especially  marked  in  the
ultrarelativistic  case  with  the   dynamic   parameter

\begin{equation}
\chi ^{2} = {e^{2}(pFFp) \over m^{6}}   \label{eq:DP}
\end{equation}

\noindent being not small even for  a  relatively  weak  field
($F_{\mu \nu }$ is  the
external field tensor,  $p_{\alpha }$ is the 4-momentum,
$m$ is the mass of the
particle). This is due to  the  fact  that  in  the  relativistic
particle rest frame the field  may  turn  out  of  order  of  the
critical one or even higher, appearing very close to the constant
crossed field.  Thus, the calculation  in constant crossed field
($\vec {\cal E} \perp \vec B$, ${\cal E}=B$)  is  relativistic  limit
of  the  calculation  in  an
arbitrary  weak  smooth  field,  possesses  a  great  extent   of
generality and acquires interest by  itself.  We  note  that,  as
($FF=F\tilde{F}=0$)  in crossed field, the dynamic parameter
$\chi ^{2}$  (\ref{eq:DP})  is
the single field invariant, by which  the  decay  probability  is
expressed. Furthermore, the calculation in crossed field is  the
least cumbersome and, therefore, we consider this case first just
to outline the calculation technique.

In the lowest order of the  perturbation  theory,  a  matrix
element of the radiative decay of  the  massive  neutrino
$\nu _{i} \rightarrow \nu _{j}\gamma $  ($i \neq j$)
in the Feynman gauge is described by the diagrams,  represented
in Fig.1 (a,b)  where double lines imply the influence of the  external
field in the propagators of intermediate particles. Summation  is
made over the virtual lepton $\ell $ in the loop ($\ell = e, \mu, \tau$).
Under  the
conditions $m^{2}_{\ell }/m^{2}_{W}\ll  1$,  $e F / m^{2}_{W} \ll 1$
the     field     induced
contribution $\Delta {\cal M}^{(F)}= {\cal M}-{\cal M}^{(0)}$  to  the
decay  amplitude  can   be
calculated in the local limit, in which the lines $W$ and $\varphi $ are
contacted to a point. It is most easily seen,  if $\Delta {\cal M}^{(F)}$
is expanded into a series in terms of the  external  field, as is shown
in Fig 2, where  the  dotted  lines  designate  the  external
electromagnetic field $A^{ex}$. We note that the first seven  diagrams
in Fig 2  coincide  with  the  diagrams  describing  Compton-like
process $\nu _{i}\gamma ^\ast \rightarrow \nu _{j} \gamma^\ast$
and, as was pointed  out  in  \cite{KuM},
this  process
amplitude is  reduced  to  the  contribution  of  the  first  two
diagrams in  the  local  limit. With  the  orthogonality  of  the
mixing matrix $K_{ij}$  taken  into  account,  this  is  due to the
fact,  that  the  main  contribution  to  the  integral
over momentum in the  loop  gives  from  the  region  of  the
virtual momenta $p \sim m_{\ell } \ll  m_{W}$. We remind that we are
investigating flavour violating processes ($i\neq j$), and, hence,
$<A>=0$, if $A$ is independent of $m_{\ell }$ . Thus, the dominant
contribution  of
order $ 1 / m^{2}_{W} \sim G_{F}$ only comes from the diagrams with
one $W$-propagator in Fig.2.

Even such a simple analysis shows the following:

\noindent 1) the predominant contribution to the amplitude is
made by the diagram ($a$) in Fig.1.
This diagram  in the local limit  of $W$-boson
propagator being contacted to a point, transforms to the diagram
shown in Fig.3;

\noindent 2) since in calculating $\Delta {\cal M}$ the  mass  of
the $W$-boson  in  the
local limit  appears in the  weak  interaction  constant
$G _{F}=g^{2}/8m^{2}_{W}$ only,   the  amplitude  does  not  contain
the   known   GIM suppression
factor of the decay $\nu _{i}\rightarrow \nu _{j}\gamma $ in vacuum
$\sim m^{2}_{\ell } / m^{2}_{W} \ll  1$ (see  Eq.(\ref{eq:M0})).

The expression for the  amplitude,  corresponding  to  the
diagram in Fig.3, can be represented in the following form:

\begin{eqnarray}
\label{eq:Am1}
\Delta {\cal M} & = & {ieG_{F}\over \sqrt 2} \; j_{\beta }
\epsilon ^{*}_{\alpha }(q) <J_{\alpha \beta }(q)> - {\cal M}^{(0)}, \\
\label{eq:Int1}
J_{\alpha \beta }(q) & = & \int d^{4}X Sp\left[ \gamma _{\alpha }
\hat{S}(X) \gamma _{\beta } (1+\gamma _{5}) \hat{S}(-X) \right] e^{iqX},
\end{eqnarray}

\noindent where $X=x-y$ and $\hat{S}(X)$ is the propagator of a
charged  lepton  in  the
crossed field (see Appendix A, Eqs.(A.1) and (A.2)).
All the other quantities in
(\ref{eq:Am1}) are defined above (see Eqs.(\ref{eq:DM})-(\ref{eq:M0})).
The details of the
tensor $J_{\alpha \beta }(q)$ calculation may be found in Appendix A,
while here we only give the result of the calculation:

\begin{eqnarray}
\label{eq:Am2}
\Delta {\cal M} & = & {e G_{F} \over 4 \pi^2 \sqrt 2}
\bigg < e (\tilde{F}f^{*}) \,
{(qFFj) \over (qFFq)} \, I_1 + {e \over 8 m^2_\ell} \,
(F \tilde{f}^{*})(qj) \, I_{2}  \nonumber \\
& + & {e^2 \over 24 m^4_\ell} \, (F \tilde{f}^{*}) \, (q\tilde{F}j) \,
I_3 + {e^2 \over 48 m^4_\ell} \, (Ff^{*}) \, (qFj) \, I_4 \bigg >, \\
f_{\alpha \beta} & = & q_\alpha \epsilon_\beta - q_\beta \epsilon_\alpha,
\nonumber \\
\tilde{f}_{\mu \nu } & = & {1 \over 2} \epsilon_{\mu \nu \alpha \beta} \,
f_{\alpha \beta}, \nonumber
\end{eqnarray}

\noindent where $F_{\mu \nu}$,
$\tilde{F}_{\mu \nu} = \epsilon_{\mu \nu \alpha \beta} \,
F_{\alpha \beta} / 2$
are  the tensor  and the dual tensor of a constant field;
$e > 0$  is  the  elementary  charge,  $G_F$  is  the  Fermi
constant. In Eq.(\ref{eq:Am2}) $I_a$ ($a = 1, \ldots, 4$)
are integrals  of  the  known Hardy-Stokes functions $f(u)$:

\begin{eqnarray}
I_1 & = & \int^{1}_{0} dt \, uf(u),  \nonumber \\
\label{eq:Int2}
I_2 & = & \int^{1}_{0} dt \, (1 - t^2) \, uf(u), \\
I_3 & = & \int^{1}_{0} dt \, (1 - t^2) \, (3 - t^2) \, u^2 \,
{df \over du}, \nonumber \\
I_4 & = & \int^{1}_{0} dt \, (1 - t^2) \, (3 + t^2) \, u^2 \,
{df \over du}, \nonumber \\
\label{eq:HS}
f(u) & = & i \; \int^{\infty }_{0} dz \exp \left [ -i \,
(zu + {1 \over 3} z^3) \right ], \\*[.5\baselineskip]
u & = &
( e^2 (qFFq) / 16 m^6_\ell )^{- 1/3} \, (1 - t^2)^{- 2/3} \nonumber
\end{eqnarray}

\noindent As can be readily checked, the amplitude (\ref{eq:Am2})
is evidently gauge invariant,
as it is expressed in terms  of  the  tensors  of  the
external field $F_{\mu \nu}$ and the photon field $f_{\mu \nu}$.
In the  weak  field limit
($F/F_e \ll 1$) the predominant contribution to the amplitude  is
made by the second term in the eåpression (\ref{eq:Am2}).
The integrals $I_a$ can be easily evaluated in  this limit,
taking  into  account
the orthogonality of the lepton mixing matrix $K_{i \ell}$~:

\begin{eqnarray}
< I_1 > & = & < 1 + O [(F/F_e)^2] > \; \sim \; O [(F/F_e)^2],
\nonumber \\
I_2 & \simeq & 2 / 3 ,
\label{eq:EInt2} \\
I_3 & \simeq & 28 / 15 , \nonumber \\
I_4 & \sim & O [(F/F_e)^4].    \nonumber
\end{eqnarray}

\noindent As expected, the amplitude (\ref{eq:Am2})
in view of (\ref{eq:EInt2})
in the weak field limit coincides with the expression (\ref{eq:DM}).
The amplitude  of  the process $\nu_i \rightarrow \nu_j \gamma$
in  a  crossed  field  (\ref{eq:Am2})
is simplified substantially in two cases:
that of the decay of a neutrino at  rest ($E_\nu = m_\nu$) and
that of the decay of  an  ultrarelativistic  neutrino ($E_\nu \gg m_\nu$).

\subsection{Neutrino at rest ($E_\nu = m_\nu$)}

In this case, the dynamic parameter (\ref{eq:DP})

\begin{displaymath}
\chi^2_\ell = {e^2 (p_1 F F p_1) \over m^6_\ell} =
\left ( {m_\nu \over m_\ell} \; {e F \over m^2_\ell} \right )^2
\end{displaymath}

\noindent is obviously small even when
the field strength exceeds the critical values
($F \ge  m^2_\ell / e$, $m_\nu \ll m_\ell$, $\chi_\ell \ll  1$).
The decay amplitude (\ref{eq:Am2}), (\ref{eq:Int2})
can then be reduced to the form:

\begin{eqnarray}
\Delta {\cal M} & \simeq & {e G_{F} \over 60 \pi^2 \sqrt 2} \,
\bigg \lbrace ({\cal F} \tilde{f}^*) \left [ (j {\cal F} {\cal F} q) +
{5 \over 4} (jq) - {7 \over 6} (q \tilde{\cal F} j) \right ]
\nonumber \\
& - & {19 \over 24} ({\cal F} f^*) (q {\cal F} j) \bigg \rbrace
(K_{ie} K^*_{je}).   \label{eq:Am3}
\end{eqnarray}

Here we have introduced the dimensionless  field tensor
${\cal F}_{\mu \nu} = F_{\mu \nu} / F_e$, where
$F_e = m^2_e / e$ is the critical strength value,
$m_e$  is   the electron  mass.
It  is  clear  from Eq.(\ref{eq:Am3})  that  the  decay
probability is represented by a polynomial  of the sixth  degree  in
field strength. In  the  limit $F \ll  F_e$ (a weak  field  limit),  the
expression for the decay probability is determined  by  the  lowest
power of $F$  and has the following form:

\begin{equation}
w_{weak} \simeq  {\alpha \over 18 \pi } \; {G^2_F \over 192 \pi^3} \;
m^5_i \, \left ( 1 - {m^2_j \over m^2_i} \right )^5 \,
\left ( {F \over F_e} \right )^2 \,
| K_{ie} K^*_{je} |^2,      \label{eq:Ww}
\end{equation}

\noindent where $m_i$ and $m_j$ are the masses of the initial
and final neutrinos.
In the opposite case, $F \gg F_e$ (a strong field limit),
we have:

\begin{equation}
w_{st} \simeq  {\alpha \over 4 \pi} \; {G^2_F \over (15 \pi)^3} \;
m^5_i \, \left ( 1 - {m^2_j \over m^2_i} \right )^5 \,
\left ( 1 + 5 {m^2_j \over m^2_i} \right ) \,
\left ( {F \over F_e} \right )^6
| K_{ie} K^*_{je} |^2.       \label{eq:Ws}
\end{equation}

\noindent The expressions~(\ref{eq:Ww}) and~(\ref{eq:Ws})
should be compared with  the  well
known probability of the decay $\nu_i \rightarrow \nu_j \gamma$
in vacuum~\cite{Pet}:

\begin{equation}
w_0 \simeq  {27 \alpha \over 32 \pi} \; {G^2_F \over 192 \pi^3} \;
m^5_i \, \left ( {m_\tau \over m_W} \right )^4 \,
\left ( 1 + {m^2_j \over m^2_i} \right ) \,
\left ( 1 - {m^2_j \over m^2_i} \right )^3 \,
| K_{i\tau} K^*_{j\tau} |^2.    \label{eq:W0}
\end{equation}

\noindent The comparison demonstrates the catalyzing effect
of the external crossed field on the decay probabilities,
as there is  no
suppression $\sim (m^2_\ell / m^2_W)$ in Eqs.(\ref{eq:Ww})
and (\ref{eq:Ws}). Actually, the enhancement influence of the external
field takes place in the weak field limit ($F \ll F_e$) under the
condition: $ F > 2 \cdot 10^{-3} F_e$.
Besides, the decay in the strong crossed field (\ref{eq:Ws}) is catalyzed
by an additional factor of the form $\sim (F / F_e)^6 \gg 1$.

\subsection{Ultrarelativistic neutrino ($E_\nu \gg m_\nu $)}

Notice that in the ultrarelativistic limit the kinematics  of
the decay $\nu_i (p_1) \rightarrow \nu_j (p_2) + \gamma (q)$ is such
that the momentum  4-vectors of the initial neutrino $p_1$
and the decay products $p_2$  and $q$  are
almost parallel to each other. Therefore, the current 4-vector
$j_\alpha = \bar \nu_j (p_2) \gamma_\alpha (1 + \gamma_5) \nu_i (p_1)$
is also proportional to these vectors
$(j_\alpha \sim p_{1 \alpha} \sim q_\alpha \sim p_{2 \alpha}$).
In this case,
the expression for amplitude (\ref{eq:Am2}) can be simplified
and reduced to the form

\begin{equation}
\Delta {\cal M} \simeq  {e^2 G_F \over \pi^2} \;
(\epsilon^* \tilde F p_1) \;
\left [ (1-x) + {m^2_j \over m^2_i} (1+x)
\right ]^{\mbox{\normalsize $1 \over 2$}} \;
< I_1 >,      \label{eq:Am4}
\end{equation}

\noindent where $x=\cos \vartheta$, $\vartheta$ is the angle between
the vectors $\vec p_1$ (the momentum of the decaying ultrarelativistic
neutrino) and $\vec {q'}$ (the photon  momentum  in  the  decaying
neutrino $\nu_i$ rest frame). The argument $u$ of the Hardy-Stokes
function $f(u)$  in  the integral $I_1$ (see Eq.(\ref{eq:Int2}))
in the ultrarelativistic case has the form:

\begin{equation}
u = 4 \left [ (1+x) (1-t^2)
\left ( 1 - {m^2_j \over m^2_i} \right ) \,
\chi_\ell \right ]^{\mbox{\normalsize - ${2 \over 3}$}}.
\label{eq:Uu}
\end{equation}

The Lorentz-invariant decay probability $w E_\nu$ can be
expressed in terms of the integral of the
squared amplitude over the variable $x$:

\begin{eqnarray}
w E_\nu & \simeq & {1 \over 16 \pi} \,
\left ( 1 - {m^2_j \over m^2_i} \right ) \;
\int\limits^{+1}_{-1} dx \; | \Delta {\cal M} |^2
\label{eq:WEe} \\
& = & {\alpha \over 4 \pi} \; {G^2_F \over \pi^3} m^6_e \chi^2_e \,
\left ( 1 - {m^2_j \over m^2_i} \right )
\int\limits^{+1}_{-1} dx \left [ (1-x) + {m^2_j \over m^2_i} (1+x)
\right ] \, | < I_1> |^2.
\nonumber
\end{eqnarray}

\noindent For small values of the dynamic parameter ($\chi_\ell \ll 1$),
the integral $I_1 (\chi_\ell)$ is expanded into the following series:

\begin{eqnarray}
I_1 & \simeq & 1 + {1 \over 15} \tilde{\chi}^2_\ell +
{4 \over 63} \tilde{\chi}^4_\ell + \ldots,
\nonumber \\
\tilde{\chi}_\ell & = & {1+x \over 2} \;
\left ( 1 - {m^2_j \over m^2_i} \right ) \; \chi_\ell,
\label{eq:Int3}
\end{eqnarray}

\noindent and the probability~(\ref{eq:WEe}) can be represented
in the form:

\begin{equation}
w \simeq {\alpha \over 4 \pi} \; {G^2_F \over (15 \pi)^3} \;
\frac{m^6_e}{E_\nu} \;
\chi^6_e \;
\left ( 1 - {m^2_j \over m^2_i} \right )  \,
\left ( 1 + 5 {m^2_j \over m^2_i} \right )  \,
| K_{ie} K^*_{je} |^2.
\label{eq:WEs}
\end{equation}

\noindent
As the dynamic parameter
$\chi_\ell \sim (E_\nu / m_\ell) (F / F_\ell)$ is proportional to the
neutrino's energy, it is clear from Eq.~(\ref{eq:WEs})
that, with increasing the energy of the decaying neutrino, the decay
probability increases as $\sim E^5_\nu$.

For great values of $\chi_\ell \gg 1$, using the asymptotic
behavior of the Hardy-Stokes function both at large and at small values
of the argument and also the unitarity of the mixing matrix $K_{i \ell}$,
one can represent Eq.(\ref{eq:WEe}) in the form:

\begin{eqnarray}
w & \simeq & {\alpha \over 4 \pi} \; {G^2_F \over \pi^3} \;
\frac{m^6_e}{E_\nu} \;
\chi^2_e \; \left ( 1 - {m^4_j \over m^4_i} \right ) \;
\Bigg \lbrace
{\mbox{\normalsize $| K_{ie} K^*_{je} |^2$, $\chi_e \gg 1$,
$\chi_{\mu, \tau} \ll 1$,}
\atop
\mbox{\normalsize $| K_{i \tau} K^*_{j \tau} |^2$,
$\chi_e \gg \chi_\mu \gg 1$, $\chi_\tau \ll 1$,}}
\label{eq:WEL1}
\end{eqnarray}

\noindent
In this way, the decay probability increases
linearly $\sim E_\nu$ ($\chi_e \gg 1$, $\chi_\tau \ll 1$)
and finally ($\chi_\tau \gg 1$) becomes a constant:

\begin{eqnarray}
w & \simeq & {21.7 \alpha \over \pi} \; {G^2_F \over \pi^3} \;
\frac{m^6_\tau}{E_\nu} \;
\chi_\tau \; | K_{i \tau} K^*_{j \tau} |^2,
\label{eq:WEL2}
\\
& & \chi_\tau \, \left ( 1- {m^2_j \over m^2_i} \right ) \gg  1.
\nonumber
\end{eqnarray}

\noindent
Comparing the decay probabilities
(\ref{eq:WEs})--(\ref{eq:WEL2})  in the crossed field
with the vacuum decay  probability  (\ref{eq:W0}), we notice that the
catalyzing effect of the field on the ultrarelativistic  neutrino
decay becomes even more substantial  compared  to  the  situation
with the neutrino at rest,  because  there  is  no  suppression
caused by the smallness of the mass of the decaying neutrino.
Recall   also   that none of the eåpressions for the
decay probability in the crossed field contain the well-known suppression
GIM-factor $(m_\ell / m_W)^4$  characteristic of the  probability
of the decay $\nu_i \rightarrow \nu_j \gamma$ in  vacuum.
The probability of neutrino decay at rest in a strong crossed field
(see Eq.(\ref{eq:Ws})) is enhanced by the additional factor
$\sim (F / F_e)^6 \gg 1$.

   Here we estimate a number of gamma-quanta which can be
result as the decay product from a neutrino beam of a high
energy accelerator in a volume filled with an electromagnetic
field. As the experimentally accessible strengths of
electromagnetic fields are
significantly below the critical strength ($F \ll F_e$)
in the laboratory conditions we can
use the expression~(\ref{eq:WEL1}) for the decay probability of
a high energy neutrino ($E_\nu \gg m_e F/F_e$, $\chi_e \gg 1$).

The number of gamma-quanta  $\Delta N^{\gamma}$
which are produced in a volume filled with a magnetic field
of the strength $B$ transversal to the neutrino beam can be
presented in the following form:

\begin{equation}
 \Delta N^{\gamma} \sim 10^7 \, \left ( \frac{B}{B_e} \right )^2 \,
 \left ( \frac{L}{1 \, m} \right ) \,
 \left ( \frac{W}{10^{19} \, GeV} \right ) \,
| K_{i e} K^*_{j e} |^2,
\label{eq:DeltaN}
\end{equation}

\noindent where $L$ is a longitudinal dimension of the volume. The parameter
$W$ has a simple meaning of a full energy of neutrinos
passed through a ``target'' during the experiment. It can be
presented as

\begin{displaymath}
 W = \int E \, \frac{dN^\nu}{dE} \, dE,
\end{displaymath}

\noindent where $dN^\nu/dE$ is the energy distribution of the
neutrino beam.
One can estimate this parameter from a number $N^{cc}$ of
$\nu_\mu's$ charge current (CC) events in detector with known
design parameters during the time of neutrino experiment.

With the expected data on $N^{cc}$ from CERN-SPS neutrino beam
which is presented, for example in ref.~\cite{Gomes}, one can estimate:

\begin{displaymath}
 W  \sim  10^{19} \, GeV.
\end{displaymath}

\noindent As one can see from~(\ref{eq:DeltaN}),  $\Delta N^{\gamma} \sim 1$
requires $B \sim 10^{10} \, G$ in a volume $\sim 1 \, m^3$.

\section{Constant Magnetic Field}

The probability of the massive neutrino decay
$\nu_i \rightarrow \nu_j \gamma$
in a constant magnetic field having the strength $\vec B$
is described by two invariant parameters:

\noindent 1) the above-mentioned dynamical parameter

\begin{equation}
\chi^2_\ell = {e^2 (p_1 F F p_1) \over m^6_\ell} =
{B^2 \over B^2_\ell} {p^2_{1 \perp} \over m_\ell^2},  \label{eq:XP}
\end{equation}

\noindent where $B_\ell = m^2_\ell / e$ is the critical magnitude of the
magnetic  field, $p_{1 \perp}$ is the initial neutrino's momentum
component, normal  to  the magnetic field;

\noindent 2) the external magnetic field intensity parameter

\begin{equation}
\eta^2_\ell = - {e^2 (FF) \over 2 m^4_\ell} = {B^2 \over B^2_\ell},
\label{eq:EP}
\end{equation}

\noindent where $F_{\mu \nu}$ is the constant uniform magnetic field
tensor, $(FF) = -2 B^2$.
Since magnetic fields in up-to-date superconductive magnets range
up to strengths $B \le 10^{5} \, G$, the hope to obtain marked quantum
effects induced by a magnetic  field,  as  it  seems,  should  be
related only to the ultrarelativistic neutrino  decay,  when  the
dynamic parameter $\chi_\ell$  may  be  not  small, while the intensity
parameter $\eta_e \ll 1$. In this case we have the crossed  field limit
which was considered in detail in the preceding section. The decay in a
strong magnetic field ($B \ge B_e = 4.41 \cdot 10^{13} \, G$)
is probably of interest in astrophysics or in early Universe cosmology.
By mentioning astrophysics,we, first of all, mean intensive magnetic
fields,  ``frozen  in''  neutron  star  substance.  The  primordial
magnetic fields, ``frozen in'' the cosmological plasma  could  also
have had the strenghts $B \ge B_e$  at  some  stages  of  the  early
Universe evolution \cite{V}.

The amplitude of the decay $\nu_i \rightarrow \nu_j \gamma$ in a magnetic
field  is
described by the same effective Feinman diagram as that shown  in
Fig.3.  The  expression  for  the  lepton  propagator   and   the
calculation details are given in Appendix~A, here we only  present
the external-field-induced contribution $\Delta {\cal M}^{(B)}$ to
the decay amplitude.

\begin{eqnarray}
\Delta {\cal M}^{(B)} & \simeq & \frac{e}{(4 \pi)^2} \,
\frac{G_F}{\sqrt 2} \Big [ i (f \tilde{\varphi}) (qj) <Y_1> +
(f \tilde{\varphi}) (q \tilde{\varphi} j) <Y_2> \nonumber \\
& + & (f \varphi) (q \varphi j) <Y_3> +
i (f \tilde{\varphi}) (q \Lambda j) <Y_4> \Big ].
\label{eq:DMB}
\end{eqnarray}

\noindent We recall that here
$j_\mu = \bar{\nu_j} (p_2) \gamma_\mu (1 + \gamma_5) \nu_i (p_1)$
is the neutrino current,

%\begin{displaymath}
\begin{equation}
< Y (m_\ell) > = \sum_{\ell = e, \mu, \tau} K_{i \ell} K^*_{j \ell} \,
Y (m_\ell),
%\end{displaymath}
\end{equation}

\noindent $p_1$, $p_2$, $q$ are the 4-momenta of the initial and final
neutrino  and the photon, respectively,
$f_{\alpha \beta} = q_\alpha \epsilon_\beta - q_\beta \epsilon_\alpha$
is the  photon  field tensor.
We  have  introduced  the  dimensionless  tensors  of  the
external magnetic field $\varphi_{\alpha \beta} = F_{\alpha \beta} / B$,
$\tilde{\varphi}_{\alpha \beta} = \epsilon_{\alpha \beta \mu \nu}
\varphi_{\mu \nu} / 2$,
$\Lambda_{\alpha\beta} = (\varphi \varphi)_{\alpha\beta}$.
The double integrals $Y_{i}$ ($i$ = 1, 2, 3, 4), entering (\ref{eq:DMB}),
have the form:

\begin{eqnarray}
Y_i & = & \int\limits_{0}^{1} dt \, \int\limits_{0}^{\infty}
\frac{dz}{\sin z} \,e^{\mbox{\normalsize $i \Omega (z, t)$}} y_i (z, t),
\nonumber \\
y_1 & = & (1 - t^2) \sin z + t \sin{zt} -
\frac{1 - \cos z \cos{zt}}{\sin z} ,
\nonumber \\
y_2 & = & \cos{zt} - (1 - t^2) \cos z - \frac{t \sin{zt}}{\tan z} ,
\label{eq:YI} \\
y_3 & = & - \cos{zt} + \frac{t \sin{zt}}{\tan z} +
2 \frac{\cos{zt} - \cos z}{\sin^2 z} ,
\nonumber \\
y_4 & = & 2 y_1 - (1 - t^2) \sin z ,
\nonumber \\
\Omega (z, t) & = & \frac{1}{2} \, \left ( \frac{z (1 - t^2)}{2} +
\frac{1}{\tan z} - \frac{\cos{zt}}{sin z} \right ) \,
\frac{(q \Lambda q)}{e B} - \frac{z}{\eta_\ell}.
\nonumber
\end{eqnarray}

\noindent It should be noted that in (\ref{eq:YI}) the integration with
respect to $z$ is performed over the complex $z$ plane. The integrand's
singularities are bypassed in a usual way \cite{S} in  the  lower
semi-plane ($\mbox{Im } z < 0$) of the complex $z$ plane.
In the weak field
approximation $B \ll B_\ell$ ($\eta_\ell \ll 1$) the integrands
in~(\ref{eq:YI})  become
rapidly oscillating functions of the variable $z$ (since at $z \simeq 1$
we have $\Omega (z,t) \simeq 1 / \eta_\ell \gg 1$).
Therefore,  the  predominant contribution to the integrals
comes from a narrow region of small $z \simeq \eta_\ell$,
in which the integrals can be  considerably  simplified.
By  straightforward  calculation,  it  can  be  shown  that   the
amplitude~(\ref{eq:DMB}) in this limit ($\eta_\ell = B / B_\ell \ll 1$,
however $\chi_\ell = (B / B_\ell) (p_\perp / m_\ell)$  has  not
to  be  small), as  should  be  expected,
coincides with the amplitude of the neutrino radiative  decay  in
the  crossed  field~(\ref{eq:Am2}).  Note, that  this  is  the  necessary
(but, certainly, not  sufficient)  condition  for  our   combersome
calculations of the magnetic-field-induced amplitude
$\Delta {\cal M}^{(B)}$ to  be correct.

In the strong field approximation $B \gg  B_\ell$ ($\eta_\ell \gg 1$), to
evaluate the integrals~(\ref{eq:YI}), it  is  convenient  to  rotate  the
integration loop clockwise by $\pi / 2$ in  the  complex $z$-plane.  In
this  case  the  amplitude
$\Delta {\cal M}^{(B)}$~(\ref{eq:DMB}),~(\ref{eq:YI})  is  significantly
simplified and can be represented as follows:

\begin{equation}
\Delta {\cal M}^{(B)}_{st} \simeq {e \over 24 \pi^2} \,
{G_F \over \sqrt 2} \, (f \tilde \varphi)
\Big [ (jq) + (j \tilde{\varphi} q) + (j \Lambda q) \Big ]
< \eta_\ell H (4 m^2_\ell / q^2_{\perp} >,
\label{eq:DMS}
\end{equation}

\noindent where $q^2_\perp = (q \Lambda q)$ is the square of
the photon momentum, which  is normal to the magnetic field,

\begin{eqnarray}
H(x) & = & {3 \over 2} \, x \left (
{x \over \sqrt{x - 1}} \arctan{\frac{1}{\sqrt{x -1}}} - 1
\right ) \qquad x > 1,
\label{eq:HX} \\
H(x) & = & - {3 \over 4} \, x  \left [
{x \over \sqrt{1-x}}
\ln \left ( {1 + \sqrt{1-x} \over 1 - \sqrt{1-x}} \right )
+ 2 + i \pi {x \over \sqrt{1-x}} \right ]
\quad x < 1. \nonumber
\end{eqnarray}

Note that, if $q^2_{\perp} = 4 m^2_\ell$ ($x=1$),
the amplitude~(\ref{eq:DMS})  has  a  root
singularity associated with the known  root  singularity  of  the
probability of the $e^+ e^-$  pair  generation  by  a  photon  in  an
external   magnetic field~\cite{Kl}.   Such   a   behavior   of   the
decay amplitude in the vicinity of $q^2_\perp \to 4 m^2_\ell$ exactly
corresponds to the singularity of the imaginary part of the photon
polarization operator in a magnetic field~\cite{BKS}.

On the basis of the expression~(\ref{eq:DMS}) for  the  amplitude  we
obtained the following expression  for  the  probability  of  the
ultrarelativistic neutrino of a moderate-energy
($p^2_\perp < 4 m^2_\ell$) in  a  strong  magnetic  field
($\eta_\ell \gg 1$):

\begin{eqnarray}
w_{st} & \simeq & {2 \alpha \over \pi} \,
{G^2_F \over \pi^3} \, \frac{m_e^6}{E_\nu} \,
\left ( \frac{B}{B_e} \right )^2 \,
| K_{i e} K^*_{j e} |^2 \, J (z),
\label{eq:W2} \\
J (z) & = & \int \limits_0^z \, dy \, (z - y) \,
\left [ \frac{1}{y \sqrt{1 - y^2}} \,
\arctan \frac{y}{\sqrt{1 - y^2}} - 1 \right ]^2,
\nonumber \\
z & = & \frac{E_\nu \sin \theta}{2 m_e}.
\nonumber
\end{eqnarray}

\noindent
To  estimate
the enhancement of the  decay  by  the  external  magnetic  field
numerically, it is sufficient to compare the
expression~(\ref{eq:W2})
for the decay probability in  a  magnetic  field  with  the
expression~(\ref{eq:W0})
for the probability of the decay of the massive
neutrino in vacuum.

As we can see,
in the relativistic neutrino decay probability~(\ref{eq:W2})
there is no suppression caused by the smallness of  the
neutrino mass, because the probability is  virtually  independent
of the neutrino mass (if $m^2_j / m^2_i \ll 1$).
Finally,  the  decay  in  a
strong field is also catalyzed by a factor $\sim (B / B_e)^2$.
Especially impressive is the comparison of the  moderate-energy
$(E^2_\nu < 4 m^2_e)$
relativistic neutrino's lifetime $\tau^{(B)}$ in the radiative decay
$\nu_i \rightarrow \nu_j \gamma$ in a strong magnetic field ($B \gg B_e$)

\begin{equation}
\tau^{(B)} \simeq {2 \cdot 10^7 \over | K^*_{je} K_{ie} |^{2}} \,
\left ( {B_e \over B} \right )^2
\left ( {1 \, MeV \over E_\nu} \right ) \; \sec
\label{eq:TB}
\end{equation}

\noindent and the lifetime $\tau^{(0)}$ in the radiative decay in vacuum

\begin{equation}
\tau^{(0)} \simeq {10^{50} \over | K^*_{j \tau} K_{i \tau} |^2} \,
\left ( {1 \, eV \over m_\nu} \right )^6
\left ( {E_\nu \over 1 \, MeV} \right ) \; \sec.
\label{eq:T0}
\end{equation}

\noindent So, the magnetic catalysis of the massive neutrino radiative
decay might solve the problem of  whether  a  heavy  enough
($m_\nu \ge 20 \, eV$) neutrino exists in the Universe.
In fact, if  the  magnetic
field fluctuations in an early enough ($t \ge 1 \, sec$) Universe  would,
for some reasons, reach such values as $B \sim 10^3 B_e$,  the  massive
neutrino's lifetime under such fields could, according to~(\ref{eq:TB}),
reduce to values of order of one second.

\section{Plane monochromatic wave}

The  development  of  intensive   electromagnetic  field
generation techniques and the current possibility  to  obtain
waves of high strength of electromagnetic field, namely
${\cal E} \sim 10^9 \, V/cm$,
stimulate the  investigation  of  quantum  processes  in  strong
external wave fields.
Indeed, the  so called wave intensity parameter

\begin{equation}
\mbox{\ae}^2_\ell = - {e^2 a^2 \over m^2_\ell},
\label{eq:XiP}
\end{equation}

\noindent (where $a$ is the amplitude of wave, $m_l$ is the
lepton mass and $e$ is an elementary charge),
characterizing the effect of the electromagnetic wave
should not be neglected.
It's  also  worth  noting  that   the   energy-momentum
conservation law for the radiative decay      in  the  wave  also
contains, along with the 4-momenta $p_1$, $p_2$, $q$, the
4-wavevector~$k$.
The process amplitude is calculated by standard Feynman rules,
in which  for  the  propagators  of  intermediate  fermions  (see
Appendix~B) exact solutions are used of  the  corresponding  wave
equations. In this section we consider
the external field of a  monochromatic  circularly  polarized
wave with the 4-potential

\begin{equation}
A_\mu = a_{1 \mu} \cos \varphi + \xi a_{2 \mu} \sin \varphi, \qquad
\varphi = kx,
\label{eq:AP}
\end{equation}

\noindent where $k^\mu = (\omega, \vec k)$ is the 4-wavevector,
$k^2 = (a_1 k) = (a_2 k) = (a_1 a_2)~=~0$, $a^2_1 = a^2_2 = a^2$;
the parameter $\xi = \pm 1$  indicates  the  direction  of  the
circular polarization (left- or rightward). Note that vectors $\vec a_1$,
$\vec a_2$ and $\vec k$ form a right-handed coordinate system.
The $S$-matrix element of the given process, just as previously,
can be represented in the following form:

\begin{equation}
S = S_0 + \Delta S,
\label{eq:SF}
\end{equation}

\noindent where $S_0$ is the matrix element of the radiative
decay of the massive neutrino in a vacuum and
$\Delta S$ is the contribution  induced by the wave field,

\begin{equation}
\Delta S = {i (2 \pi)^4 \over
\sqrt{2 E_1 V \cdot 2 E_2 V \cdot 2 q_0 V}} \,
\sum^{+2}_{n=-2} {\cal M}^{(n)} \delta^{(4)} (nk + p_1 - p_2 - q)
\label{eq:DS}
\end{equation}

\noindent Here $p_1$, $p_2$, $q$ and $E_1$, $E_2$, $q_0$ are the 4-momenta
and the energies of the initial, final neutrinos and photon, respectively,
$n = 0, \pm 1,\pm 2$ is the difference between the numbers of absorbed
and  emitted photons of the wave field.

Note, that the matrix element of some process in  the  field
of an electro\-mag\-ne\-tic wave has usually the form of summation of $n$
type~(\ref{eq:DS}), where $-\infty < n < \infty$ \cite{BLP}.
That only five values of $n$  in
our case  are  possible  is  extraordinary  and  is  due  to  the
following reasons. The process $\nu_i \rightarrow \nu_j \gamma$
is local with the  typical scale $\Delta x \le 1/m_f$
($m_f$ is the mass of the  virtual  fermion).  In
this case the  angular  momentum  conservation  degenerates  to  spin
conservation. Since the total spin of the particles participating
in this process is no greater than 2, $| n |_{\max} = 2$  is  the  maximum
difference between the numbers of absorbed and emitted photons of
the external field (the photons of  a  monochromatic  circularly
polarized  wave  have  a  definite  spin $\xi = \pm 1$).  The  direct
calculation supports this conclusion. A  similar  phenomenon  has
been discovered before~\cite{BTV} in  studies  of  the  effect  of  a
circularly polarized wave on flavour --  changing  transitions  of
the massive neutrinos $\nu_i \leftrightarrow \nu_j$ ($i \neq j$)
with $| n |_{\max} = 1$.

Notice that in the uniform constant fields the decay
$\nu_i \rightarrow \nu_j \gamma$
with $m_i > m_j$ is valid only. This is due to the fact that the
energy-momentum conservation law in this fields coincides with the one
in vacuum. On the other  hand,  as  it  follows  from Eq.(\ref{eq:DS}),
the external electromagnetic wave field  can  also  induce
radiative transition with $m_i \le m_j$  forbidden  without  the  field.
Indeed, from the energy-momentum conservation  law  in  the  wave
field

\begin{displaymath}
nk + p_1 = p_2 + q
\end{displaymath}

\noindent the relation follows

\begin{displaymath}
m^2_i - m^2_j \ge  - 2 n (k p_1)
\end{displaymath}

\noindent In such a manner, the radiation  decay
$\nu_i \rightarrow \nu_j \gamma$
with $m_i \le m_j$  is possible on condition that $n > 0$.

In the lowest order of perturbation theory, a matrix element
of the radiative transition $\nu_i \rightarrow \nu_j \gamma$
is described by the  effective
diagram represented in Fig.3, in which not only $W$-boson, but  also
(for $i=j$) $Z$-boson is exchanged at the point $x$. The expressions
for the fermion propagator in the wave's field, as well  as  some
details of the awkward calculation of the invariant amplitudes
${\cal M}^{(n)}$ are given in Appendix~B. Here we only present
the final result:

\begin{eqnarray}
{\cal M}^{(0)} & = & - {e \over 16 \pi^2} \, {G_F \over \sqrt 2} \,
\frac{4}{(k q)}
\nonumber \\
& \times & \Big \lbrace \sum_\ell
( K_{i \ell} K^*_{j \ell} + {1 \over 2} \delta_{i j} g_\ell ) \,
\mbox{\ae}^2_\ell m^2_\ell
\left [ (jfk) J_1 (m_\ell) + (j \tilde f k) J_2 (m_\ell) \right ]
\nonumber \\
& - & {3 \over 2} \delta_{ij} \sum^{}_{q} Q_q g_q \mbox{\ae}^2_q m^2_q
\left [ (jfk) J_1 (m_q) + (j \tilde f k) J_2 (m_{q}) \right ]
\Big \rbrace,
\nonumber \\
{\cal M}^{(\sigma)} & = & - {e \over 16 \pi^2} \, {G_F \over \sqrt 2} \,
\frac{e \sigma (\tilde f F^\sigma)}{(kq)}
\Big \lbrace
\sum^{}_\ell ( K_{i \ell} K^*_{j \ell} - {1 \over 2} \delta_{ij} )
\label{eq:Am5} \\
& \times & \left [ ( (p_1 - p_2) j) J_3^{(\sigma)} (m_\ell) +
8 {m^2_\ell \mbox{\ae}^2_\ell \over (kq)} \, (jk) \,J^{(\sigma)}_4
(m_\ell) \right ]
\nonumber \\
& + & {3 \over 2} \delta_{ij} \sum^{}_{q} (2 T_{3q} Q^2_q)
\left [ ((p_1 - p_2) j) J^{(\sigma)}_3 (m_q) +
8 {m^2_q \mbox{\ae}^2_q \over (kq)} (jk) J^{(\sigma)}_4 (m_q) \right ]
\Big \rbrace,
\nonumber \\
{\cal M}^{(2 \sigma)} & = & - {e \over 16 \pi^2} \, {G_F \over \sqrt 2} \,
\frac{e^2 (f F^\sigma) (j F^\sigma q)}{(kq)^2}
\nonumber \\
& \times & \Big \lbrace
\sum^{}_\ell ( K_{i \ell} K^*_{j \ell} + {1 \over 2} \delta_{ij} g_\ell )
J^{(\sigma)}_5 (m_\ell)
- {3 \over 2} \delta_{ij} \sum^{}_{q} (Q^3_q g_q) J^{(\sigma)}_5 (m_q)
\Big \rbrace,
\nonumber \\
F^{(\sigma)}_{\mu \nu} & = & k_\mu a^{(\sigma)}_\nu -
k_\nu a^{(\sigma)}_\mu,
\qquad
a^{(\sigma)}_\mu = (a_1 + i \xi \sigma a_2)_\mu,
\nonumber \\
g_f & = & 2 T_{3f} - 4 Q_f \sin^2 \theta_w, \qquad f = \ell, q.
\nonumber
\end{eqnarray}

\noindent Here $\sigma$ is the sign of the summation index $n$ in
Eq.~(\ref{eq:DS}) $(\sigma  = \pm 1)$,
the index $\ell$ indicates charged leptons ($\ell = e, \mu, \tau$) and
the index $q$ indicates quark flavors ($q = u, c, t, d, s, b$), $T_{3f}$
is the third component of the weak isospin and $Q_f$ is the electric
charge in units of the elementary charge, $m_\ell$ and $m_q$ are the
masses of the virtual leptons and quarks, $K_{ij}$ is the unitary lepton
mixing matrix,

\begin{eqnarray}
J_1 (m_f) & = & \int\limits^{1}_{0} dy \int\limits^{\infty }_{0} d\tau
\left [ \tau
\left ( j^2_1 + {3 + y^2 \over 1 - y^2} j^2_0 \right ) -
{4 y^2 \over 1 - y^2} j_0 j_1 \right ]
e^{\mbox{\normalsize $i \Phi (m_f)$}},
\nonumber \\
J_2 (m_f) & = & \int\limits^{1}_{0} dy {1 + y^2 \over 1 - y^2} \,
\int\limits^{\infty }_{0} d\tau \, \tau j_0 j_1
e^{\mbox{\normalsize $i \Phi (m_f)$}},
\nonumber \\
J^{(\sigma)}_3 (m_f) & = & \int\limits^{1}_{0} dy
\int\limits^{\infty }_{0} d\tau ( j_0 - i \sigma j_1 )
e^{\mbox{\normalsize $i (\Phi (m_f) - \sigma \tau)$}},
\label{eq:Int4} \\
J^{(\sigma)}_4 (m_f) & = & \int\limits^{1}_{0} {dy \over 1 - y^2}
\int\limits^{\infty }_{0} d\tau \, \tau^2 j_0 ( j^2_0 + j^2_1 )
e^{\mbox{\normalsize $i (\Phi (m_f) - \sigma \tau)$}},
\nonumber \\
J^{(\sigma)}_5 (m_f) & = & \int\limits^{1}_{0} dy
\int\limits^{\infty}_{0} d\tau \, \tau ( j^2_0 + j^2_1 )
e^{\mbox{\normalsize $i (\Phi (m_f) - 2 \sigma \tau)$}},
\nonumber \\
\Phi (m_f) & = & - {4 \tau \over 1 - y^2} \, {m^2_f \over (kq)} \,
\left [ 1 + \mbox{\ae}^2_f ( 1 - j^2_0 ) \right ],
\nonumber \\
\mbox{\ae}_f^2 & = & - Q_f^2 \, {e^2 a^2 \over m^2_f}.
\nonumber
\end{eqnarray}

\noindent Here $j_0 = \sin \tau / \tau$, $j_1 = - d j_0 / d\tau$
are   so   called   Bessel   spherical harmonics;
the  other  denotations  and  quantities   in~(\ref{eq:Am5})
and~(\ref{eq:Int4})
have been introduced above. It  is  easy  to  see  that  the
amplitudes ${\cal M}^{(n)}$ are explicitly gauge-invariant and do not
contain divergences. Note that the expressions we  have  obtained
for the amplitudes ${\cal M}^{(n)}$ of the radiative transition
$\nu_i \rightarrow \nu_j \gamma$
in the field of a monochromatic wave allow a simple check.
Indeed, if in Eqs.~(\ref{eq:Am5}) and~(\ref{eq:Int4})
the wave frequency $\omega$ tends to zero (i.e. $k_\mu \to 0$),
the strengths $\vec {\cal E}$ and $\vec B$ of the electric and magnetic
fields being fixed, then in this limit the amplitude of the decay
$\nu_i \rightarrow \nu_j \gamma$
in crossed field must be obtained
(see Eqs.~(\ref{eq:Am2})--(\ref{eq:HS})):

\begin{equation}
\Delta {\cal M} [ \mbox{Eq.~(\ref{eq:Am2})} ] = \sum^{+2}_{n=-2}
{\cal M}^{(n)}
\Bigg \vert_{\mbox{\normalsize $k_\mu \to (0, \vec 0)$}
\atop \mbox{\normalsize ${\cal E}$, B -- fix}} .
\label{eq:Am6}
\end{equation}

\noindent To prove the conclusion~(\ref{eq:Am6}),
it is necessary to take into account that:

\noindent 1) in the above limit, the tensor $F^{(\sigma)}_{\mu \nu}$
is expressed in  terms  of the strength tensor
$F_{\mu \nu}$  and  dual  tensor $\tilde {F}_{\mu \nu}$
of the crossed field:

\begin{equation}
F^{(\sigma)}_{\mu \nu}
\Bigg \vert_{\mbox{\normalsize $k \to 0$}
\atop \mbox{\normalsize ${\cal E}$, B -- fix}}
= i \sigma ( F_{\mu \nu} + i \sigma \xi \tilde{F}_{\mu \nu}),
\label{eq:FCF}
\end{equation}

\noindent 2) if we take advantage of the identity

\begin{eqnarray}
(A_1 \tilde A_2)_{\alpha \beta} + (A_2 \tilde A_1)_{\alpha \beta}
= {1 \over 2} (A_1 \tilde A_2) g_{\alpha \beta},
\nonumber \\
(A_1 \tilde A_2)_{\alpha \beta} = A_{1 \alpha \rho}
\tilde A_{2 \rho \beta},
\qquad
(A_1 \tilde A_2) = A_{1 \alpha \rho} \tilde A_{2 \rho \alpha},
\nonumber
\end{eqnarray}

\noindent where $A_{1 \mu \nu}$, $A_{2 \mu \nu}$ are arbitrary
antisymmetric  4-tensors,
then the wave inten\-si\-ty parameter $\mbox{\ae}_\ell^2$~(\ref{eq:XiP})
can be related to the  dynamic parameter
$\chi_\ell^2$~(\ref{eq:DP}) by the expression:

\begin{equation}
\left [ \mbox{\ae}^2_\ell \left ( {p_1 k \over m^2_\ell} \right )^2
\right ] \Bigg \vert_{\mbox{\normalsize $k \to 0$}} = \chi^2_\ell,
\label{eq:CDX}
\end{equation}

\noindent 3) to remove the undeterminacy arising at $k \to 0$ in the
expressions~(\ref{eq:Am5}) for ${\cal M}^{(n)}$ it is useful to take
advantage of the limit relationship:

\begin{equation}
{ k_\alpha \over (ka)} \Bigg \vert_{\mbox{\normalsize $k \to 0$}} =
{(bFF)_\alpha \over (bFFa)},
\label{eq:LK}
\end{equation}

\noindent where $a$ and $b$ are arbitrary 4-vectors
(however, $b$ is a timelike vector),

\noindent 4) in the crossed field limit the terms in $M^{(n)}$
(\ref{eq:Am5}) proportional to $\delta_{ij}$ do not make any contribution
(because of kinematics reasons).

\noindent In passing to the limit using~(\ref{eq:FCF})--(\ref{eq:LK})
the result~(\ref{eq:Am6}) is reproduced im\-me\-dia\-te\-ly.

The probability of the transition $\nu_i \rightarrow \nu_j \gamma$
in the wave field

\begin{equation}
w = \sum^{+2}_{n=-2} w^{(n)}
\label{eq:WPW}
\end{equation}

\noindent is, in general, rather awkward. We shall give it only in
the most interesting, from the physical point of view, case of the
transition $\nu_i \rightarrow \nu_j \gamma$ with the initial neutrino
$\nu_i$ being ultrarelativistic ($E_\nu \gg m_\nu$). Despite the fact
that the wave intensity parameter $\mbox{\ae}^2_\ell$~(\ref{eq:XiP})
under laboratory  conditions  cannot be great (e.g. for laser fields
$\omega \sim 1 \, eV$, ${\cal E} \le 10^9 \, V/cm$,
$\mbox{\ae}^2_e \le 10^{-3}$),
substantial enhancement of the transition probability  is possible.
The main effect  of  the  enhancement connects with  the
decrease and even complete disappearance (for $n > 0$ as  we  shall
see below) of the suppression factor caused by  the smallness  of
the neutrino mass. Recall that an analogous result  was  obtained
for uniform and constant fields and was discussed  in  the  above
sections  (see  Eqs.~(\ref{eq:WEs})--(\ref{eq:WEL2}) and~(\ref{eq:W2})).
In the ultrarelativistic limit, the probabilities $w^{(n)}$ at $n \le 0$
remain suppressed:

\begin{eqnarray}
E_\nu w^{(-2)} & \sim & O \left ( \alpha \,
{G^2_F m^{10}_\nu \over m^4_e} \mbox{\ae}^4_e \right ),
\nonumber \\
E_\nu w^{(-1)} & \sim & O \left ( \alpha \,
{G^2_F m^8_\nu \over m^2_e} \mbox{\ae}^2_e \right )
\label{eq:EW1} \\
E_\nu w^{(0)} & \sim &
O \left ( \alpha G^2_F m^2_\nu m^4_e \mbox{\ae}^4_e \right ).
\nonumber
\end{eqnarray}

\noindent The other probabilities $w^{(n)}$ ($n$ = $+1$, $+2$)
in the limit $E_\nu \gg m_\nu$  are
substantially simplified to be represented in the following form:

\begin{eqnarray}
E_\nu w^{(+1)} & \simeq & {4 \alpha \over \pi} \, {G^2_F \over \pi^3} \,
m_e^6 \mbox{\ae}_e^6
| K_{i e} K_{j e}^* - \frac{1}{2} \delta_{i j} |^2
\nonumber \\
& \times & \int^{+1}_{-1} dx \, {1-x \over (1+x)^2} \,
| J_4^{(+1)} (m_e) |^2,
\label{eq:EW2} \\
E_\nu w^{(+2)} & \simeq & {\alpha \over 4 \pi} \,
{G^2_F \over \pi^3} \, (p_1 k) \, m_e^4 \mbox{\ae}_e^4
| K_{i e} K_{j e}^* + \frac{1}{2} \delta_{i j} g_e |^2
\nonumber \\
& \times & \int^{+1}_{-1} {dx \over 1+x}
\left [ {(1 - \xi) \over 2} + {(1 - x)^2 \over 4} \, {(1 + \xi) \over 2}
\right ] \, | J_5^{(+1)} (m_e) |^2.
\nonumber
\end{eqnarray}

\noindent Here, the integration is performed with respect to
$x=\cos \vartheta$,  $\vartheta$ being
the angle between the photon momentum $\vec q$ and the wave vector
$\vec k$ in the center-of-mass  of  the  final  neutrino $\nu_j$ and
photon $\gamma$. Consequently, in the ultrarelativistic case in the
integrals $J^{(+1)}_4$, $J^{(+1)}_5$ (see Eqs.~(\ref{eq:Int4})) the
substitution is needed $(qk) \simeq (1+x) (p_1 k) / 2$.
The comparison of the probabilities
$w^{(n)}$~(\ref{eq:EW1})--(\ref{eq:EW2})
of the transition $\nu_i \rightarrow \nu_j \gamma$ in the wave field and
the probability $w_{0}$~(\ref{eq:W0})
of the decay $\nu_i \rightarrow \nu_j \gamma$ in vacuum shows that
the  probabilities $w^{(n)}$ at $n = +1, +2$  do  not
contain  suppression  associated  with  the
smallness of the neutrino mass (recall that  the  probability  of
the decay of an ultrarelativistic neutrino in vacuum is
$w_0 \sim m^6_\nu / E_\nu$).
Below we estimate the ratio of the probability $w$ (\ref{eq:WPW}) of
the  transition $\nu_i \rightarrow \nu_j \gamma$ for a neutrino from
a high-energy accelerator in the wave field of the laser type and the
probability $w_{0}$~(\ref{eq:W0}) of the decay in vacuum:

\begin{equation}
R = {w \over w_0} \sim 10^{33} \,
\left ( {1 \, eV \over m_\nu} \right )^6
\left ( {E_\nu \omega \over m^2_e} \right )^5
\left ( 10^3 \mbox{\ae}^2_e \right )^2,
\label{eq:RPW}
\end{equation}

\noindent where the wave intensity parameter $\mbox{\ae}^2_e$
for the laser type fields is:

\begin{equation}
\mbox{\ae}^2_e \simeq  10^{-3}
\left ( {{\cal E} \over 10^9 \, V/cm} \right )^2
\left ( {1 \, eV \over \omega} \right ).
\label{eq:XiP1}
\end{equation}

\noindent Such strong enhancement of the $\nu_i \rightarrow \nu_j \gamma$
transition probability, even at relatively small wave intensity
($\mbox{\ae}^2_e \le 10^{-3}$) appears rather impressive.

The results obtained in this section may be of interest  for
astrophysics and cosmology. In particular, in the wave field  the
process of the photon splitting into the neutrino pair
$\gamma \rightarrow \nu_i \tilde{\nu_j}$ becomes possible.
This process probability has the form:

\begin{eqnarray}
w_{\mbox{\normalsize $\gamma \to \nu_i \bar \nu_j$}} & \simeq &
\frac{\alpha}{3 \pi} \,
\frac{G_F^2}{8 \pi^3} \, \frac{m_e^4}{q_0} \, \mbox{\ae}_e^4
\Big \lbrace 8 m_e^2 \mbox{\ae}_e^2 | K_{i e} K_{j e}^* -
\frac{1}{2} \delta_{i j} |^2 \, | J^{(+1)}_4 (m_e) |^2
\nonumber \\
& + & (qk) | K_{i e} K_{j e}^* + \frac{1}{2} \delta_{i j} |^2 \,
| J_5^{(+1)} (m_e)|^2 \Big \rbrace .
\label{eq:WG}
\end{eqnarray}

\noindent As is easily seen from~(\ref{eq:WG}), this process probability,
in the same way, is not suppressed by the smallness of the neutrino mass.
It can be treated as an additional mechanism of the energy loss by stars.

\section{Conclusion}

In this work, in the framework of the  Standard  Model  with
fermion  mixing,  we have investigated  the  effect  on  the  process
$\nu_i \rightarrow \nu_j \gamma$ of  the  massive  neutrino  radiative
decay  of  external electromagnetic  fields  of  various configurations:
constant crossed field (section~2), constant uniform  magnetic   field
(section~3), plane monochromatic wave's field (section~4).  The
analysis of the decay amplitudes and probabilities obtained leads
to the following conclusion, which is the same for all the  field
configurations  covered:  an   external   electromagnetic   field
catalyses the massive neutrino  radiative  decay.  An  especially
strong enhancement occurs in the case  of  the  ultrarelativistic
neutrino  radiative  decay,  since  in  this   case   the   decay
probability does not contain suppression caused by the  smallness
of the neutrino's mass.

    In section~2  (see eq.~(\ref{eq:DeltaN})) we have estimated a number of
gamma-quanta which could be resulted as the neutrino decay in a
volume filled with a magnetic field. Let us give here the estimation
in the case of limiting in the laboratory conditions values of
$B$ and $W$ ($B \sim 10^9 \,G$, $W \sim 10^{19} \,GeV$):

\begin{displaymath}
 \Delta N^{\gamma} \sim 10^{-2} \,
 \left ( \frac{B}{10^9 \, G} \right )^2 \,
 \left ( \frac{L}{1 \, m} \right ) \,
| K_{i e} K^*_{j e}|^2.
\end{displaymath}

It is worth noting that the estimation we have presented is
numerically small and seems likely that there is no possibility
to carry out such neutrino experiment in the near future.

Nevertheless, the mensioned above mechanism of the electromagnetic
catalysis of the massive neutrino radiative decay is of interest
in astrophysics where gigantic neutrino fluxes and strong magnetic
fields can take place simultaneously (a process of a coalescence of
neutron stars~\cite{MEZ}, an explosion of a supernova of the type
SN 1987A~\cite{NAD}).
Let us estimate a relative flux of gamma-quanta which traverses
a domain filled with a strong magnetic field ($B \gg B_e$):

\begin{displaymath}
\frac{\Phi^\gamma}{\Phi^\nu} \sim 10^{-12} \,
\left ( \frac{B}{B_e} \right )^2 \,
\left ( \frac{L}{10 \, km} \right ) \,
| K_{i e} K^*_{j e} |^2,
\end{displaymath}

\noindent where  $\Phi^\nu$  is the neutrino flux with the average energy
$E_\nu \sim 1 \, MeV$ transversal to the magnetic field strength,
$L$ is the characteristic dimension of the domain.
Gamma-quanta produced from the neutrino decay can be observed
in astrophysical experiments provided that the domains with such strong
magnetic fields exist.

On the other hand, the results presented in the section~3 are,
in our opinion, of interest
in the cosmology of the early Universe. Indeed, in the
recent papers it was pointed out a possibility of the  generation
of primordial strong magnetic fields through thermal fluctuation
in the primordial plasma with the magnetic field strengths of order of
$10^{12} \div 10^{15} G$ ~\cite{LEM} or $10^{13} \div 10^{18} G$ ~\cite{TCS}
and coherence lengths of order of $10 \div 100 \, cm$.
Let us estimate the neutrino lifetime in the case of existence
of primordial small scale magnetic field strengths of order of
$\sim 10^{17} \div 10^{15} \, G$.
For this purpose we use the expression~(\ref{eq:TB}) we have
obtained for the moderate energy neutrino ($E_\nu \sim k T \sim 1 \, MeV$)
lifetime $\tau^{(B)}$ in a strong magnetic field ($B \gg B_e$) and
get the following estimation: $\tau^{(B)} \sim 0.1 \div 100 \, sec$.
This may be of interest in connection with the cosmological problem,
conserning the contradiction between the COBE data on the cosmic
microwave background anisotropies and the observed power spectrum
of the large-scale structure~\cite{WGS}.

\bigskip

\newpage

{\large\bf Acknowledgments}

\bigskip

The authors thank L.B.~Okun, V.A.~Rubakov, K.A.~Ter-Martirosyan
and M.I.~Vysotsky for many fruitful discussions and for helpful
remarks.
This work was supported in part by Grant N RO 3300 from International
Science Foundation and Russian Government.
The work of N.V.~Mikheev was supported by a Grant N d104 by International
Soros Science Education Program.
The work of L.A.~Vassilevskaya has been made possible by a
fellowship of INTAS Grant 93-2492-ext and is carried out within the
research program of International Center for Fundamental Physics
in Moscow.

\appendix

\section{$J_{\alpha \beta}$ calculation in constant \newline
electromagnetic field}

The amplitude corresponding  to  the  diagram  in  Fig.~3  is
calculated according to the conventional Feinman rules. In  doing
so, for the propagators of intermediate charged leptons
exact solutions are used of the corresponding wave  equations  in
the constant electromagnetic field. With the crossed field,  the
propagator of the charged lepton $\hat S^{(F)} (x, y)$
in the proper time formalism~\cite{S} has the form:

\begin{eqnarray}
\hat S^{(F)} (x, y) & = & e ^{\mbox{\normalsize $i \Phi (x,y)$}}
\hat S (X),
\label{eq:SFA} \\[3mm]
\hat S (X) \quad & = & - {i \over 16 \pi^2} \int\limits_0^\infty
{ds \over s^2} \bigg [ {1 \over 2s} (X \gamma) +
{i e \over 2} (X \tilde F \gamma) \gamma_5
\nonumber \\
& - & {s e^2 \over 3} (X F F \gamma)
 +  m - {s m e \over 2} (\gamma F \gamma) \bigg ]
\label{eq:S1} \\
& \times &  \exp \left ( - i \left [ m^2 s + {1 \over 4 s }X^2  +
{s e^2 \over 12} (XFFX)  \right ] \right ),
\nonumber
\end{eqnarray}

\noindent where $X_\mu = (x-y)_\mu$, $F_{\mu \nu}$, $\tilde F_{\mu \nu}$
are the field tensor  and  field  dual
tensor, $e > 0$ is  the  elementary  charge, $\gamma_\mu$, $\gamma_5$
are  Dirac $\gamma$-matrices (the metric, the convensional representation
of  Dirac $\gamma$-matrices, etc. correspond to the book~\cite{BLP}),
$m$ is the mass of the charge lepton, the phase $\Phi (x,y)$ is
determined in  the following way:

\begin{eqnarray}
\Phi (x,y) & = & e \,
\int\limits^{\mbox{\normalsize $x$}}_{\mbox{\normalsize $y$}}
d\xi_\mu \, K_\mu (\xi),
\nonumber \\
K_\mu (\xi) & = & A_\mu (\xi) + {1 \over 2} F_{\mu \nu} (\xi - y)_\nu.
\label{eq:FKA}
\end{eqnarray}

Owing to $ \partial_\mu K_\nu - \partial_\nu K_\mu = 0 $, the
path of integration from $y$ to $x$ in ~(\ref{eq:FKA}) is
arbitrary and, therefore,

\begin{eqnarray}
\Phi (x,\,y) + \Phi (y,\,x) = 0.
\label{eq:LF}
%\nonumber
\end{eqnarray}

\noindent Using~(\ref{eq:LF}), the integration of $J_{\alpha \beta}$
with respect to $x$ and $y$ (see Fig.~3) can be easily redused to an
integration  with  respect  to $X = x - y$ (see Eq.~(\ref{eq:Int1})).
From~(\ref{eq:Int1}) and~(\ref{eq:S1}) it is  clear  that
the integrals with respect to $X$ are Gaussian, so that they can  be
readily calculated:

\begin{eqnarray}
G & = & \int d^4X \,
e^{\mbox{\normalsize $- i \left ( \frac{1}{4} XRX + qX \right )$}}
= - (4 \pi )^2 (\det R)^{- 1/2}
e^{\mbox{\normalsize $(i q R^{-1} q)$}},
\nonumber \\
G_\mu & = & \int d^4X \, X_\mu \,
e^{\mbox{\normalsize $- i \left ( {1 \over 4} XRX + qX \right )$}}
= i {\partial G \over \partial q^\mu},
\label{eq:SG} \\
G_{\mu \nu} & = & \int d^4X \, X_\mu X_\nu
e^{\mbox{\normalsize $- i \left ( {1 \over 4} XRX + qX \right )$}}
= - {\partial^2 G \over \partial q^\mu \partial q^\nu}.
\nonumber
\end{eqnarray}

\noindent In the remaining double integral with respect to the proper
times $s_1$, $s_2$, it is convenient to pass to  the  dimensionless
variables $z$, $t$:

\begin{eqnarray}
z = m^2 (s_1 + s_2), & & t = {s_1 - s_2 \over s_1 + s_2}, \qquad
ds_1 ds_2 = {1 \over 2 m^4} z dz dt,
\nonumber \\
0 \le z \le \infty, & & -1 \le t \le 1,
\label{eq:NV}
\end{eqnarray}

\noindent The substitution into the amplitude~(\ref{eq:Am1}) of the
expression  for $J_{\alpha \beta}$ in the form of a double
integral with respect to $z$, $t$ results
in the final expression~(\ref{eq:Am2}) and~(\ref{eq:Int2}).

In the case of a  constant  uniform  magnetic  field $\vec B$ the
propagator of the charged lepton $\hat S^{(B)} (x,y)$ in the proper
time formalism has the form:

\begin{eqnarray}
\hat S^{(B)} (x,y) & = & - {i \beta \over 2(4 \pi)^2} \,
e^{\mbox{\normalsize $i \Phi (x,y)$}}
\int\limits_0^\infty {ds \over s \, \sin (\beta s)}
\nonumber \\
& \times & \bigg \lbrace {1 \over s} \bigg [
\cos (\beta s) (X \tilde\Lambda \gamma) +
i \sin (\beta s) (X \tilde \varphi \gamma) \gamma_5
\bigg ]
\nonumber \\
& - & {\beta \over \sin (\beta s)} (X \Lambda \gamma)
%\nonumber \\
%
%& + &
+ m \left [
2 \cos (\beta s)
- \sin (\beta s) (\gamma \varphi \gamma)
\right ] \bigg \rbrace
\label{eq:SB} \\
& \times & \exp \left ( - i \left [ m^2 s + {X \tilde\Lambda X \over 4 s}
- {\beta  \over 4 \tan (\beta s)}
(X \Lambda X) \right ] \right ),
\nonumber
\end{eqnarray}

\noindent where $\varphi_{\mu \nu} = F_{\mu \nu} / B$,
$\tilde{\varphi}_{\mu \nu} = \tilde{F}_{\mu \nu} / B$
are  the  dimensionless  field
tensor and dual field tensor  of  the  constant  magnetic  field,
$\Lambda_{\alpha \beta} = (\varphi \varphi)_{\alpha \beta}$,
$\tilde\Lambda_{\alpha \beta} = (\tilde\varphi
\tilde\varphi)_{\alpha \beta}$,
$\beta = e B$, $X_\mu = (x-y)_\mu$, the phase $\Phi (x,y)$ is described
in~(\ref{eq:FKA}). Note that the propagator~(\ref{eq:SB})
can be represented in a fully covariant form,
because the parameter $\beta$ in a purely magnetic field  can
be rewritten as $\beta = e B = \sqrt{- F^2 / 2}$. The calculation
procedure for the tensor $J_{\alpha \beta}$~(\ref{eq:Int1}) in the
case of a constant magnetic field, though more awkward, does not,
in principle, differ from the case of a crossed field.

\section{On the calculation of the $S$-matrix element
in the field of a monochromatic circularly \newline polarized wave}

The propagator of the charged fermion in the field of a plane
wave with a 4-potential $A_\mu = A_{\mu} (\varphi)$ of the general
form can be obtained by method, given in ~\cite{IZ}, and has the form:

\begin{eqnarray}
\hat S(x,y) & = & \int {d^4p \over (2\pi)^4}
 \left ( 1 -  {e_f \hat k \hat A' \over 2(kp)} \right)
{ \hat p + m_f \over p^2 - m^2_f }
\left ( 1 -  {e_f \hat A \hat k \over 2(kp)} \right)
\label{eq:SVAWE}\\[3mm]
& \times & \exp \bigg \{ i \bigg [  -p ( y - x ) + {1 \over (kp)}
\int\limits^{\mbox{\large $\varphi'$}}_{\mbox{\large $\varphi$}}
d\varphi \, ( e_f (pA) +{ 1\over2 } e^2_f A^2 ) \bigg ] \bigg \} ,
\nonumber
\end{eqnarray}

\noindent where  $A_\mu  = A_\mu (\varphi)$, $\varphi = kx$,
$A'_\mu  = A_\mu (\varphi')$, $\varphi' = ky$,  $k$ is the
4-wavevector ($k^2=0$), $e_f$ and $m_f$ are the charge and the mass
of fermion, respectively.

In the case of the circularly polarized wave with 4-potential
\begin{equation}
A_\mu (\varphi) = a_{1 \mu} \cos (\varphi) +
 a_{2 \mu} \sin (\varphi),
\end{equation}
where 4-vectors $a_{1 \mu}$ and $a_{2 \mu}$ are orthogonal
to the 4-wavevector $k_\mu$:
\begin{equation}
(a_{1}a_{2}) = (a_{1}k) = (a_{2}k) = 0
\end{equation}
the expression for propagator may be represented in the
following form:
\begin{eqnarray}
\hat S(x,y) & = & \int {d^4p \over (2\pi)^4}
 \left ( 1 -  {e_f \hat k \hat A' \over 2(kp)} \right)
{ \hat p + m_f \over p^2 - m^2_f }
\left ( 1 -  {e_f \hat A \hat k \over 2(kp)} \right)
\nonumber \\[3mm]
& \times & \exp \bigg \{ i \bigg [ -p ( y - x ) -
{e_f (a_{1} p) \over (kp)} ( \sin (\varphi') - \sin (\varphi) )
\label{eq:SCIRCLE} \\[3mm]
& - & {e_f (a_{2} p) \over (kp)} ( \cos (\varphi') - \cos (\varphi) ) -
{e^2_f a^2 \over 2(kp)}(\varphi' - \varphi ) \bigg ] \bigg \}.
\nonumber
\end{eqnarray}

Since the power of the exponent of the propagator~(\ref{eq:SCIRCLE})
contains nonlinear functions of coordinates
 $x$ and $y$  $(\sin(\varphi)$, $\sin(\varphi')$,
$\cos(\varphi)$ and $\cos(\varphi'))$,
it is convenient to expand the corresponding part of an exponent in the
Fourier expansion with the coefficients of expansion
been proportional to the Bessel functions ~\cite{BLP}.

Given the integration over one of the momenta in the loop
 $d^{4}q$ it is convenient to use the following basis:
\, $ p_\mu$, $ h_{\mu\nu}p_\mu$,
$\tilde h_{\mu\nu}p_\nu$, $h_{\mu\nu} h_{\nu\beta} p_\beta$,
where
\begin{equation}
h_{\mu\nu} = k_\mu a_{1\nu} - k_\nu a_{1\mu},
\qquad
\tilde h_{\mu\nu} = k_\mu a_{2\nu} - k_\nu a_{2\mu},
\end{equation}
By using the known relation
\begin{eqnarray}
J_{0} \left ( \sqrt {b^2 - 2 b c \cos \alpha + c^2} \right) =
J_0(b)J_0(c) + 2 \sum_{s = 1}^{\infty} J_{s}(b)J_{s}(c) \cos s\alpha,
%\nonumber
\end{eqnarray}
one can reduce the remaining infinite series to five terms,
which may be brought to the form~(\ref{eq:DS}).

%%%%%%%%%%%%%%%  Figures  %%%%%%%%%%%%%%%%%%%%%%%%%%%%%%%%%%%%%%%

\begin{figure}
\centerline{\epsfysize=0.9\textheight \epsffile{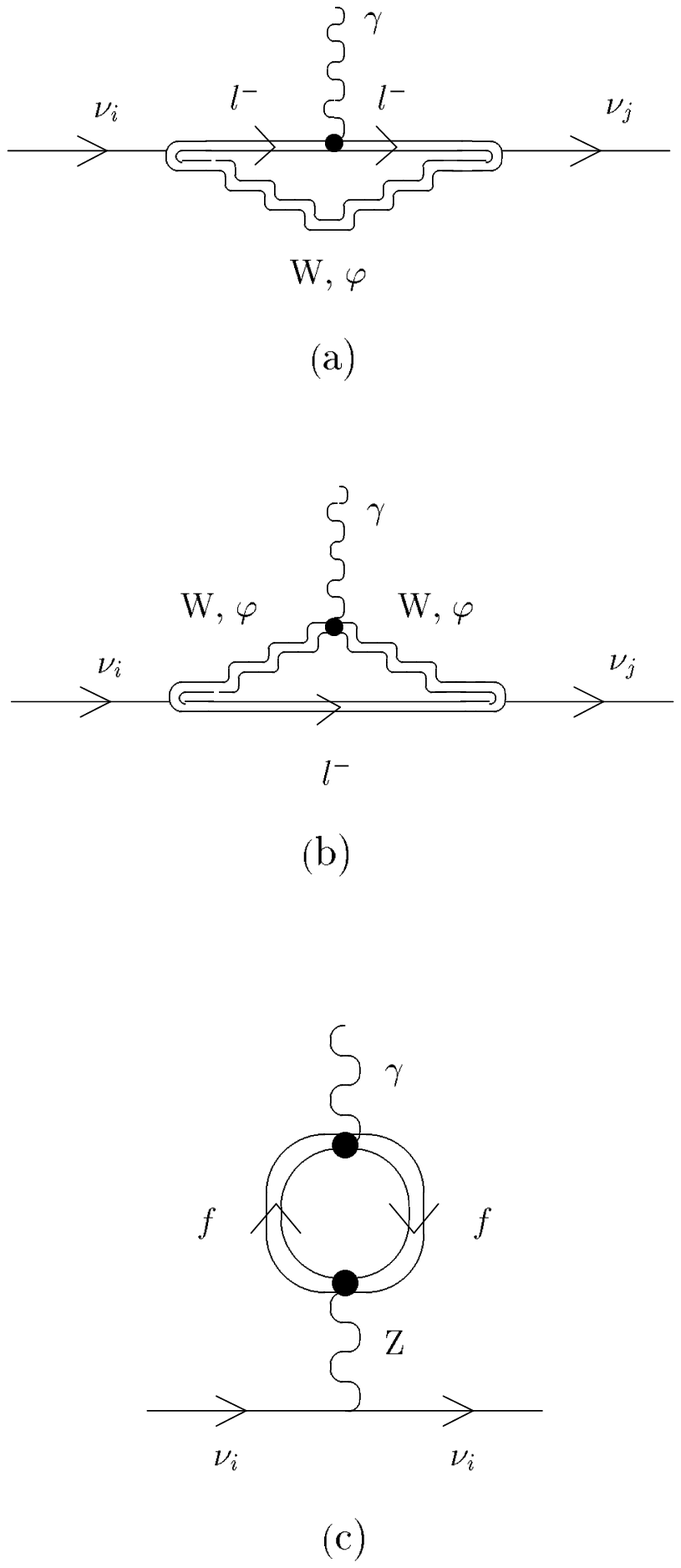}}
\caption{}
\end{figure}

\begin{figure}
\centerline{\epsfysize=0.9\textheight \epsffile{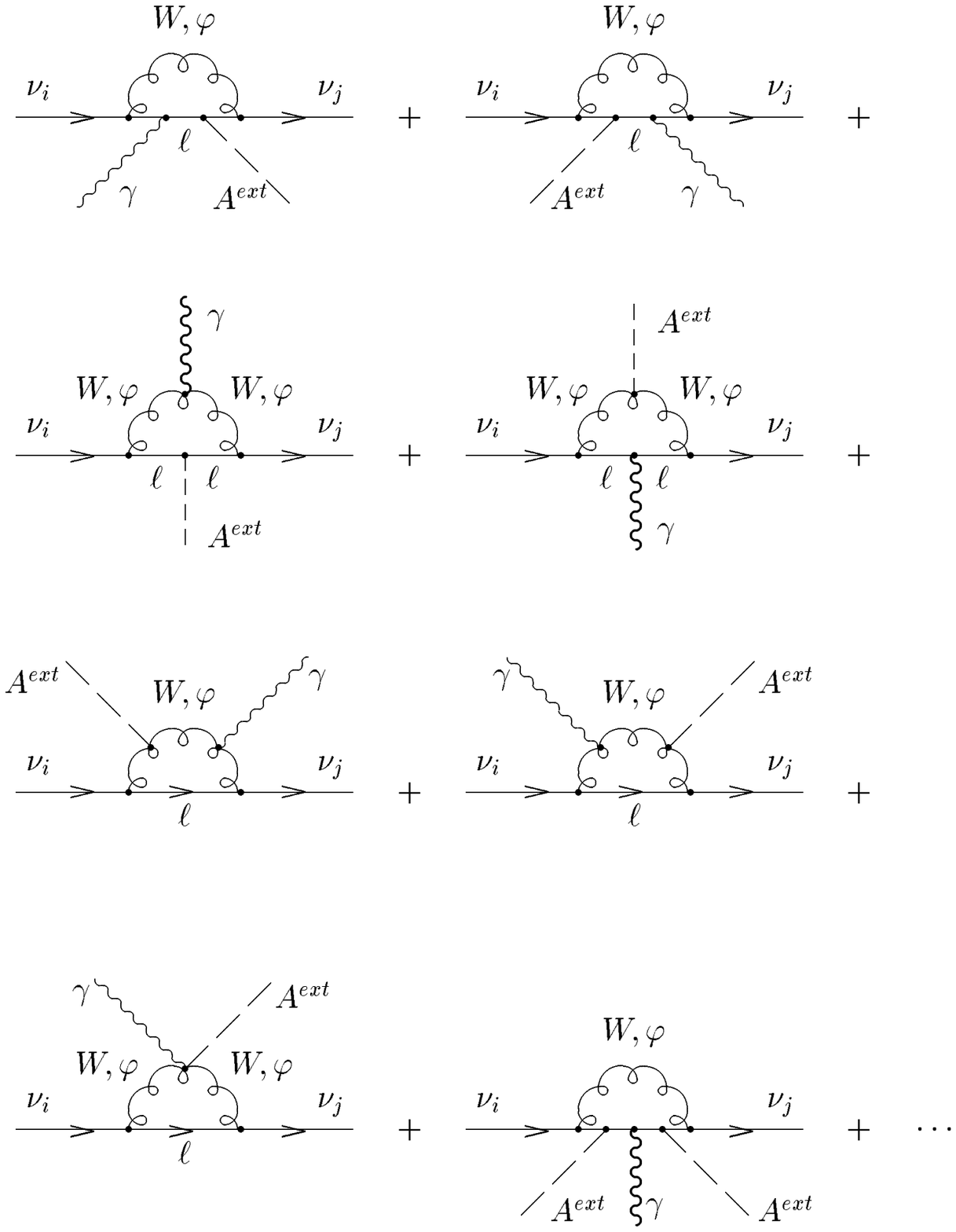}}
\caption{}
\end{figure}

\begin{figure}
\centerline{\epsfysize=0.9\textheight \epsffile{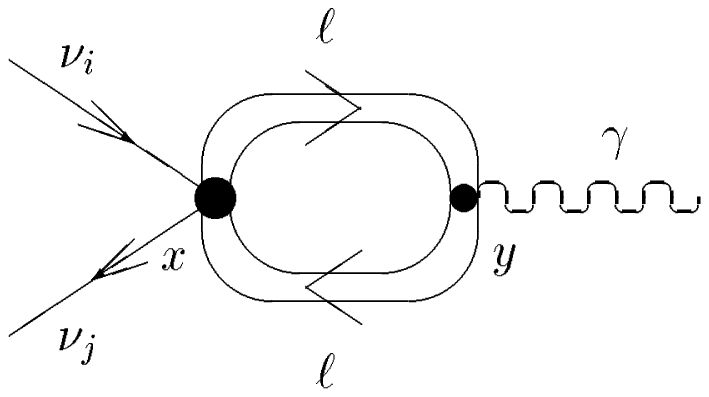}}
\caption{}
\end{figure}

%%%%%%%%%%%%%%%  end Figures  %%%%%%%%%%%%%%%%%%%%%%%%%%%%%%%%%%%%

\end{document}